# CONGESTION CONTROL FOR P2P LIVE STREAMING

Nikolaos Efthymiopoulos[1], Athanasios Christakidis[1], Maria Efthymiopoulou[1], Loris Corazza[1], Spyros Denazis[1]

[1]Department of Electrical and Computer Engineering, University of Patras, Patras, Greece

## ABSTRACT

*In recent years, research efforts tried to exploit peer-to-peer (P2P) systems in order to provide Live Streaming (LS) and Video-on-Demand (VoD) services. Most of these research efforts focus on the development of distributed P2P block schedulers for content exchange among the participating peers and on the characteristics of the overlay graph (P2P overlay) that interconnects the set of these peers. Currently, researchers try to combine peer-to-peer systems with cloud infrastructures. They developed monitoring and control architectures that use resources from the cloud in order to enhance QoS and achieve an attractive trade-off between stability and low cost operation. However, there is a lack of research effort on the congestion control of these systems and the existing congestion control architectures are not suitable for P2P live streaming traffic (small sequential non persistent traffic towards multiple network locations). This paper proposes a P2P live streaming traffic aware congestion control protocol that: i) is capable to manage sequential traffic heading to multiple network destinations , ii) efficiently exploits the available bandwidth, iii) accurately measures the idle peer resources, iv) avoids network congestion, and v) is friendly to traditional TCP generated traffic. The proposed P2P congestion control has been implemented, tested and evaluated through a series of real experiments powered across the BonFIRE infrastructure.*

## 1. INTRODUCTION

Video streaming has become a dominant part of today's internet traffic. As analysed in [7] between 2012 and 2013, the highest growth happened on the Internet side in online video with 16 per cent year-over-year growth. By 2018, digital TV and online video will be the two most highly penetrated services, with 86 per cent and 78 per cent respectively. Additionally is considered that the market is growing with very good chances of very high penetration of these services to new internet users. On the other hand the tremendous number of users with heterogeneous capabilities leads even the major streaming service providers (e.g. YouTube) to suffer from high bandwidth costs. Peer-to-peer live streaming and video on demand architectures [2],[6],[25] have received a lot of research attention in the past few years aiming at achieving a better trade-off between bandwidth costs and quality of the transmitted video, while providing scalability and stability of these services. In more detail, the major requirements for P2P live streaming systems are:
Efficiency of the video distribution, as analysed in [1],[2],[3],[14], in terms of upload bandwidth utilization of participating peers. The goal here is to minimize the additional bandwidth that is contributed by a set of media servers (cloud). Efficiency has a direct impact on the trade-off between bandwidth costs and video quality.

Stability of the system, as described in [4],[5],[20],[21],[22],[23],[24], in the presence of dynamic conditions. The stability of the system is greatly affected by the dynamic conditions of the underlying network. The total P2P overlay bandwidth also changes quite frequently due to peer arrival and departures. These conditions have a serious impact in the quality of service (QoS) and consequently in the quality of experience (QoE). A stable P2P live steaming system must be able to monitor and react to these changes.

Scalability property of such systems is determined by the amount of bandwidth and processing overhead that media servers have to contribute as the number of participating peers grows. For the design of a scalable system, this overhead has to remain low even in cases that the number of participating peers is large.

A P2P overlay is a graph in which each node represents a user, and each edge that connects two nodes represents the exchange of video blocks between users. Several methods [2],[14],[3] have been proposed that try to optimize this graph in order to achieve stability and maximum exploitation of the available bandwidth of the participating peers while simultaneously they exploit network locality [2],[27]. These works assume the a priori knowledge of the dynamic upload bandwidth even in cases where peers fail to fully exploiting it. Under this observation there is a need for a P2P congestion control architecture which will be able to provide this information to P2P overlay optimization architectures in order to make their implementation in real P2P streaming systems feasible.

Currently, monitoring and control systems have been proposed [5],[8],[1],[29] that try, in a scalable and dynamic fashion, to monitor the available resources of a P2P overlay in order to calculate the deficit or surplus of its aggregate upload bandwidth. In this way, in case of deficit, they allocate dynamically additional upload bandwidth or, in case of surplus, they exploit it for other purposes. Nevertheless, these attempts, and other that explore the dynamics of P2P live streaming [26],[28] are based on the dynamic and accurate estimation of the idle upload bandwidth of each participating peer and its upload bandwidth capacity that a successful P2P congestion control architecture will offer.

Although there is a vast amount of literature, which analysed in detail in [9], on congestion control for point-to-point bulk data transfers, there are only very few works regarding P2P congestion control. An approach that concerns P2P traffic is LEBDAT [13] but it is suitable for a P2P file sharing system where: i) traffic is persistent, ii) traffic consisted from much larger blocks than those used in P2P LS and VoD , iii)there are no delay constraints in the application. Only a recent approach that described in [12] proposes a congestion control algorithm for P2P live streaming. Despite its good features it assumes persistent traffic and transmits in parallel the various video blocks to multiple receivers. In this way resources are wasted in case of no persistent traffic and the latency for the reception of a block is highly increased. Thus it doesn't achieve low delay which is an essential requirement in P2P live streaming systems.

Motivated by the lack of critical mass of research in the area of congestion control for P2P LS and VoD systems and the serious issues raised above, we have designed, implemented and evaluated in a real environment a congestion control P2P architecture that:

- Is suitable for highly dynamic traffic characterized by sequential transmissions to different network locations (P2P video blocks)
- It efficiently utilizes the upload bandwidth of participating peers
- It remains stable and robust in the eventuality of changes in peer's upload bandwidth, time-varying delays and dynamic underlying network conditions

- It accurately and dynamically measures the available upload bandwidth capacity of each peer and avoids buffer overloading of the participating network devices in the underlying network (BufferBloat problem)

The reminder of this paper is structured as follows: Section 2 presents our P2P live streaming system's architecture. Section 3 provides the problem setting. In Section 4 is analysed the proposed P2P congestion control architecture. Section 5 presents the P2P congestion control strategy. Section 6 describes our evaluation test-bed and evaluates the proposed P2P congestion control architecture. Finally in Section 7 we conclude and highlight our future steps.

## 2. SYSTEM ARCHITECTURE

Our P2P live video streaming system (Fig. 1) consists of a media server in a cloud, (noted by S) and a set of peers (noted by N). The cloud is responsible for: i) the initial diffusion of the video to a small subset of nodes among participating peers, ii) the tracking of the network addresses of participating peers in order to assist the construction and management of the P2P overlay, iii) the dynamic and scalable monitor of the resources of participating peers, iv) the dynamic allocation and release of auxiliary bandwidth.

The video stream that the system disseminates is divided into video blocks. In order to allow peers to exchange video blocks, each peer maintains network connections with a small subset of other peers which will be noted as neighbours. The sets of these connections change dynamically and form a dynamic graph called the P2P overlay. In our previous works [1],[2],[3] we present a graph topology and P2P overlay management (dynamic and distributed optimization) algorithms that each peer periodically executes which result in the dynamic reconfiguration of the P2P overlay. We use distributed optimization theory in order to dynamically ensure in a distributed (scalable) and dynamic fashion that: i) peers have connections proportional with their upload bandwidth, ii) peers have connections with other peers close to the underlying network, iii) our P2P overlay is adaptable to underlying network changes and peer arrivals and departures. This allows us to efficiently exploit all the available bandwidth resources even if they are highly heterogeneous.

The dynamic output of the P2P overlay management algorithms that run in each participating peer is a neighbour list that is passed in the Distributed Block Transmission Scheduler.

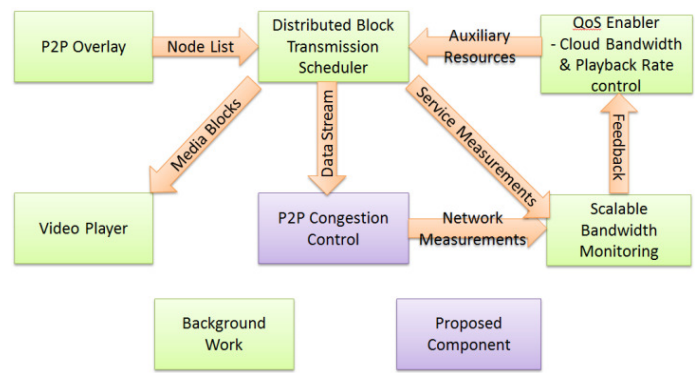

Figure 1. Proposed P2P live streaming system architecture

After that, video block exchanges are coordinated by the Distributed Media-Block Transmission Scheduler (DBTS) which is comprised by a set of algorithms executed by every peer who dynamically communicates with its neighbours. The major objective of DBTS is to ensure the timely delivery of every video block to every peer by exploiting the upload bandwidth of participating peers and the additional bandwidth resources that media servers (cloud) may contribute. Each peer periodically sends to its neighbours control messages which encapsulate information about video blocks that it owns. Thus, periodically each peer (through a matching algorithm) is able to request from each one of its neighbours a different video block or nothing if there is no video block to request. In order to perform the requests a matching algorithm is executed periodically by each peer and its objective is to request as many unique blocks as possible. These requests are served sequentially by each peer who prioritizes them by selecting each time its most deprived neighbour to serve its block request. As most deprived is defined the neighbour that has the smallest total number of blocks compared to the video blocks that sender peer owns. Our proposed DBTS is analysed in detail in our previous works [1],[2],[3]. DBTS sends the video blocks that have to be sent in the P2P congestion control component and the ordered stream with the blocks that it receives to the video player.

Our proposed P2P overlay and our DBTS enhance our P2P live streaming system with two properties. The first property (Property 1) is that if idle bandwidth exists it is derived from bandwidth surplus in the system and not from the inefficiency of the system to exploit it. In other words, we guarantee that the presence of idle bandwidth implies (testifies) the complete stream delivery. The second property is that the percentages of the idle resources among participating peers are almost equal (Property 2). We highlight here that in case of heterogeneous peer upload bandwidth various peers send with various bitrates (analog with their upload bandwidth capacity) but the percentage of their bandwidth utilization, and so the percentage of their idle time is very similar.

By exploiting the aforementioned properties we developed two components responsible for the monitoring of the total upload bandwidth of the P2P overlay and the control of the auxiliary upload bandwidth and the playback rate in order to have a stable P2P live streaming system. These are background work and we describe them in detail in [18] and [19].

We note the first as Scalable Bandwidth Monitoring (SBM) in which a scalable gossip protocol that is connected with a centralized component in the cloud and : i) aggregates the monitoring information from DBTS and P2P congestion control, ii) forms all the required metrics that QoS enabler needs.

QoS Enabler, which is the second one, has to calculate dynamically the amount of total system's upload bandwidth surplus or deficit towards the control of the idle bandwidth resources. In order to achieve this it: i) add or remove dynamically the amount of upload bandwidth that is needed as this is determined by the bandwidth allocation control strategy and/or ii) adapt the playback rate to the available resources.

Our P2P congestion control is able to manage sequential transmissions of video blocks to multiple locations that DBTS sends to it and to provide to the Scalable Bandwidth Monitoring and to the P2P overlay the dynamic estimation of: i) the upload bandwidth capacity, ii) the idle bandwidth resources of each participating peer with the way that will be requested from the latter. In the rest of this work we describe this component in detail.

## 3. PROBLEM STATEMENT

Without loss of generality, we assume that the source of congestion problems lies with the uploading capabilities of the peers rather than the downloading. Likewise, the problem will be aggravated by the incoming edge of the network (usually home gateways or DSLAM) as they may act as the primary bottleneck for any of the outgoing flows including P2P flows.

Accordingly, the goal of the congestion control method is to control the queue size of this bottleneck node by controlling the number of network packets that should be injected to the network. From the viewpoint of the P2P LS or VoD, this queue should always be non-empty, in order to fully utilize the available bandwidth provided of course that the application has the necessary blocks to transmit. In addition, its size shouldn't increase over time as this would lead to congestion problems and packet loss. The control of this queue is carried out by observing the latencies between the source sender peer and its various destinations.

In more detail as it is depicted in Fig. 2 each peer by acting as a sender sends sequentially P2P blocks (B1-B5). Each one of them is composed from a set of network packets and is heading to a different receiver peer that belongs in a set of receiver peers (Receiver Peer 1,2 and 3 in Fig. 2). The delay between the sender and each receiver peer i is different. Each receiver peer i sends acknowledgement packets. In Fig. 2 delays and acknowledgments are noted as d(i) and ack(i) respectively. Furthermore between the sender and each of the receivers there is a bottleneck network point which forwards packets with a dynamic bitrate that we note here as h(t). The objective of the proposed control strategy is to estimate h(t) and control the size of this queue by using as feedback the acknowledgements that derived from receiver peers which have different and variable delays. We highlight here that the traditional congestion control approach is not functional because of the diversity of these delays.

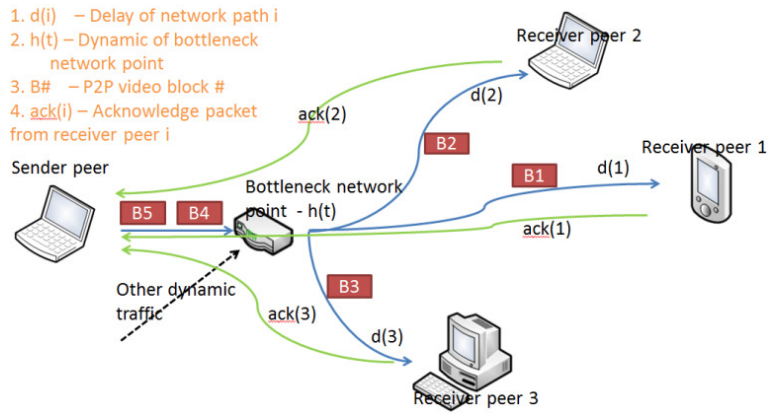

Figure 2. P2P live streaming network traffic

## 4. P2P LIVE STREAMING CONGESTION CONTROL ANALYSIS

In order to fulfil the requirements that we described earlier a control algorithm is executed periodically and ensures the stable congestion control. The proof of its stability is out of the scope of this work and analysed in detail in [9]. Our congestion control algorithm runs periodically with a period T. Each time that the algorithm is executed the source injects u(kT) packets in the network according to Eq. 1.

$$u(kT) = \gamma \left[ w - \left( \sum_{j=0}^{l-1} u(jT) - \sum_{j=0}^{l-1} ack(jT) \right) \right] Eq.1$$

Where ack(jT) is the number of packets that sender acknowledges between time instants (j-1)T and jT. Parameter $\gamma$ is a constant and its value between 0 and 1 ensures the stability of our congestion control. As we will analyse later parameter w is the upper limit in which we want to set the queue size of the bottleneck network node.

In the rest of this section we prove two lemmas with very important practical consequences upon which the architecture of the proposed control method is based.

Table 1. Notation

| Symbol | Definition |
|--------|------------|
| T | Period under which the P2P congestion control algorithm is executed |
| nT, n | Discrete time instant, number of period |
| u(n) | Number of packets that should be injected in the network by the sender during the n-th period |
| h(n) | Number of packets that forwarded by the bottleneck network point during the n-th period |
| y(n) | Number of packets that are in the queue of the bottleneck network point at time instant n |
| d(n) | Delay that queue introduces to a packet that enters to it at time instant n |
| b(n) | Estimated upload bandwidth of the sender at time instant n |
| $y_{REF}$ | Desirable number of packets in the queue |
| $u_{REF}$ | Number of packets that injected in the queue in the equilibrium state in order to maintain in length $y_{REF}$ |
| $d_{REF}$ | Desirable delay of the queue |
| s | Size in bits of each packet |
| $\lambda$ | Time interval between the time instant n that control is executed and the time instant that the last acknowledged packet sent. |
| $h_{\lambda}(n)$ | Number of packets that the bottleneck network node forwards between time instants n and n-$\lambda$ |
| $u_{\lambda}(n)$ | Number of packets that sender sent between time instants n and n-$\lambda$ |
| $\gamma$ | Eigenvalue of the controlled system |

**Lemma 1**: If the control method of Eq. 1 is applied to a P2P live streaming system then the queue length in the bottleneck network node is always upper bounded by the parameter w that can be initially set.

**Proof**: Let a sender peer send packets to m receiver peers which we order according to their delays (RTT), $d_p$, between a packet transmission to a peer p and the reception of its acknowledgement by the sender, where $d_1 \leq d_2 \leq \ldots \leq d_m$. As the control algorithm runs periodically $d_p$ can be expressed as $n_p T$ where $n_p$ is equal to the ratio $d_p/T$. Now we define as y(kT) the packet queue length in the bottleneck node at time kT. Initially we have y(0)=0 which is smaller than w. Thus, in order to prove lemma 1 with induction it suffices to prove that if lemma 1 holds for time lT, which means that y(lT)<w, then it also holds for time lT+1, which means that y((l+1)T)<w. The queue length in the bottleneck network node at time instant lT is described from Eq. 2.

$$y(lT) = \sum_{j=0}^{l-1} u(jT) - \sum_{j=0}^{l-1} h(jT) \quad Eq.2$$

In Eq. 2 u(kT) is the number of packets send by the peer during the period kT, and so the number of packets that arrived in the bottleneck node, and h(kT) is the number of packets that were served from that node during that period. According to the definition of y(nT) in Eq. 2 y((n+1)T) can be calculated recursively as:

$$y\big((l+1)T\big) = y(lT) + u(lT) - h(lT)\; Eq.3$$

If in Eq. 3 we substitute u(lT) with the proposed from Eq. 1 we have:

$$y\big((l+1)T\big) = y(lT) + \gamma\left[w - \left(\sum_{j=0}^{l-1}u(jT) - \sum_{j=0}^{l-1}ack(jT)\right)\right] - h(lt)\; Eq.4$$

The sum of the acknowledged packets from t=0 to t=l-1 is equal to the sum of the sums of the packets that have been acknowledged by each receiver peer during those periods. The sum of the acknowledgments from a peer p during that period is equal to the packets that have been served from the bottleneck network node towards p from t=0 to t=l-$n_p$-1, since the acknowledgments from the packets that were served towards p in the period t=l-$n_p$-1 to t=l-1 haven't been received yet. As described earlier $n_p$ is the ratio between $d_p$ and the period with which we execute our congestion control algorithm. According to this observation Eq. 4 is reformed as:

$$y\big((l+1)T\big) = y(lT) + \gamma\left[w - \left(\sum_{j=0}^{l-1}u(jT) - \sum_{p=1}^{m}\lambda_p\sum_{j=0}^{l-np-1}h(jT)\right)\right] - h(lT)\; Eq.5$$

In Eq. 5 m is the number of receivers and $\lambda_p$ is the number of packets send to p divided by the sum of all the packets send to all the receivers.

We can split the sum of the packets that the bottleneck network node served from j=0 until j=l-1 into two sums as depicted in Eq. 6. In this equation $\lambda^1_p$ expresses the ratio of packets that were sent to node p and were served from bottleneck network node from j=0 until j=l-$n_p$-1, which have already been acknowledged (as derived from the definition of $n_p$), and $\lambda^2_p$ expresses the ratio of packets that were sent to node p and were served from bottleneck network node from j= l-$n_p$ until j=l-1, which have not acknowledged yet.

$$\sum_{j=0}^{l-1}h(jT) = \sum_{p=1}^{m}\lambda^1_p\sum_{j=0}^{l-np-1}h(jT) + \sum_{p=1}^{m}\lambda^2_p\sum_{j=l-np}^{l-1}h(jT)\;\; Eq.6$$

Now from Eq. 5 by exploiting Eq. 6 we have:

$$y\big((l+1)T\big) = y(lT) + \gamma\left[w - \left(\sum_{j=0}^{l-1}u(jT) - \sum_{j=0}^{l-1}h(jT) + \sum_{p=1}^{m}\lambda^2_p\sum_{j=l-n_p}^{l-1}h(jT)\right)\right] - h(lT)\; Eq.7$$

By using Eq. 2 we have:

$$y\big((l+1)T\big) = y(lT) + \gamma\left[w - \left(y(lT) + \sum_{p=1}^{m}\lambda^2_p\sum_{j=l-np}^{l-1}h(jT)\right)\right] - h(lT)\; Eq.8$$

By adding and subtracting w in the second part of Eq. 8 we have:

$$y\big((l+1)T\big) = w - (1-\gamma)*\big(w - y(lT)\big) - \gamma\sum_{p=1}^{m}\lambda^2_p\sum_{j=l-np}^{l-1}h(jT) - h(lT)\; Eq.9$$

From Eq. 9 i)(1-γ) is positive because 0<γ<1 ,ii) w-y(lT) is positive according to our initial assumption and iii) h(lT) is always non negative, as it represents the packets that bottleneck network node serves. Thus, we can conclude that y((l+1)T)<w

*The practical importance of this theorem is that if the buffer dedicated to these flows in the bottleneck network node is larger than w, then packet loss will not occur.*

**Lemma 2**: If the control algorithm of Eq. 1 is applied to a P2P live streaming system and w satisfies the following inequality.

$$w > u_{max}\left(\sum_{p=1}^{m}\lambda_p^2 n_p + \frac{1}{\gamma}\right) \ Eq.10$$

Then the queue length is positive for any time instant l>$n_m$+1, where $n_m$ equals to $d_m$/T for the receiver m which exhibits the greatest latency between the sender and all the potential receivers. In the above equation $u_{max}$ is the maximum number of packets that can be served by the bottleneck node in a period T.

**Proof**: From Eq. 2 and Eq. 8 we can see that y((l=$n_m$+1)T) is positive. If we prove that if y(lT) is positive then y((l+1)T) is also positive then by induction we have proved Lemma 2.
According to Eq. 8 we have:

$$y\big((l+1)T\big) = (1-\gamma) * y\big(lT\big) + \gamma w - \gamma\sum_{p=1}^{m}\lambda_p^2 \sum_{j=l-n_p}^{l-1} h\big(jT\big) - h\big(lT\big) \ Eq.11$$

Where y((l+1)T) is larger than:

$$y\big((l+1)T\big) \geq \gamma w - \gamma\sum_{p=1}^{m}\lambda_p^2 \sum_{j=l-np}^{l-1} h\big(jT\big) - h\big(lT\big) \ Eq.12$$

But according to our model the maximum number of packets that bottleneck network point serves is $u_{max}$ so we can rewrite Eq. 12 as:

$$y\big((l+1)T\big) \geq \gamma w - \gamma\sum_{p=1}^{m}\lambda_p^2 \sum_{j=l-np}^{l-1} u_{max} - u_{max} \ Eq.13$$

Where the second part of Eq. 13 is:

$$\gamma w - \gamma\sum_{p=1}^{m}\lambda_p^2 \sum_{j=l-np}^{l-1} u_{max} - u_{max} = \gamma\left[w - u_{max}\left(\sum_{p=1}^{m}\lambda_p^2 n_p + \frac{1}{\gamma}\right)\right] \ Eq.14$$

*The practical importance of this lemma is that we are able to calculate dynamically w according to the network latencies and the ratio of unacknowledged packets from each receiver and guarantee in this way that there will not be idle bandwidth resources.*

## 5. DYNAMIC WINDOW CALCULATION

In this section we analyse how it is calculated the window of the proposed P2P congestion control. Towards this goal we will exploit Lemma 1 and Lemma 2 in order to ensure that the capacity of the bottleneck network point will not be exceeded and that all the available bandwidth will be exploited. From Lemma 2, if we choose $\gamma=1$ we have:

$$w > (u_{max}/T)\sum_{p=1}^{m}\lambda_p^2 n_p T + \frac{1}{\gamma} = U\sum_{p=1}^{m}\lambda_p^2 d_p + 1 \, Eq.15$$

As described previously, $d_p$ is the time between a packet transmission to peer p and the reception of its acknowledgement by the sender. This time interval is equal to the RTT between the sender and the receiver when the queue of the bottleneck network point is empty plus the time the packet has been in this queue ($d_p=d_q+RTT_p$). $d_q$ is 0 when the queue is empty and $d_{qmax}$ when the queue is full. From this equation we can calculate the delay in the queue $d_p=d_q+RTT_p$ and thus be able to control it to a point which belongs in the region (0,dqmax), where $d_{pmax}=d_{qmax}+RTT_p$. We note here that the calculation of $d_q$ and $d_{qmax}$ is the same for every receiver peer p and consequently independent of p. As a result we note as d(kT) the average queue delay that the packets, which were sent to the various receivers during the kth interval, experienced.

The problem that arises here is the calculation of $RTT_p$ and $d_{qmax}$. Since there is no way to accurately measure the $RTT_p$ we substitute it with $d_{pmin}$, which is the lowest delay observed for peer p. In order to calculate $d_{qmax}$ we must observe packet loss due to congestion, in which case $d_{pmax}=d_{qmax}+RTT_p$ where $d_{pmax}$ is the latency of the last packet that has been successfully transmitted. We now set in Eq. 16 $d_{ref}$ which is the queue delay in which we want to operate as a percentage a of the total queue size. As a result we have:

$$d_{ref} = d_{pmin} + a*(d_{pmax} - d_{pmin}) \, with \, 0 < \alpha < 1 \, Eq.16$$

The intuition behind this is that when the peer-to-peer flows start with an empty queue it holds that $RTT_p=d_{pmin}$, otherwise, when unrelated flows pre-exists we get an inaccurate greater value for the $RTT_p$. However, this is not a problem since the goals of the proposed congestion control algorithm are not compromised. By setting $d_{ref}$ greater than $d_{pmin}$ we guarantee that there are always available packets in the queue waiting to be transmitted and thus the available bandwidth is utilized. Also, by having $d_{ref}$ always less than $d_{pmax}$ congestion and packet loss is avoided. Packet loss will trigger the right estimation of $d_{pmax}$ and so the recalibration of the control leading to the desired behaviour.

We are now in position to dynamically determine the window size, w(kT), in each iteration of the congestion control algorithm, according to the Eq.17:

$$w(kT) = \left[U(kT)\sum_{p=1}^{m}\lambda_p^2(kT)(d\_min_p + d_{ref} + T) + 1\right] + \gamma_2\left[d_{ref} - d(kT)\right] Eq.17$$

Where U(kT) is the estimated upload bandwidth in the previous interval and $\lambda_p^2(kT)$ is the ratio of the number of packets transmit to node p and not yet acknowledged to the total number of transmitted packets in the same interval that are not yet acknowledged.

The intuition behind Eq. 17 is that the first term is derived directly from Lemma 2. The second term of Eq. 17 namely $\gamma2[d_{ref}-d(kT)]$ becomes positive and increases the window in case that $d_{ref}>d(kT)$ and in this case the queue in the bottleneck network point increases. On the contrary if $d_{ref}<d(kT)$ this second term becomes negative and the queue in the bottleneck network point decreases. In this way the queue is stabilized to the desired $d_{ref}$.

The value of $d_{ref}$, and more specifically the value of the constant α in Eq. 16, determines the aggressiveness of the proposed congestion control algorithm towards unrelated TCP traffic. If α is close to 0 then the congestion control algorithm is not aggressive at all and in case it is co-existing with TCP it gives priority to the latter. In the opposite case, where α is close to 1, the algorithm becomes very aggressive with high probability of causing starvation to other TCP flows. In our future work we will focus on the analytical correlation between the value of the $d_{ref}$ and its effect in case that our P2P congestion control co-exists with TCP.

Finally the value of U(kT) is calculated dynamically by measuring the rate of the arrival of the acknowledgments during the time interval between kT and kT-$t_c$ (i.e. the last $t_c$ seconds). If this interval is small then the congestion control can react very quickly to changes in the available upload bandwidth but in expense of its stability. This type of calculation is meaningful only when the queue in non-empty, as otherwise the estimated available bandwidth will be equal to the sending rate which will be probably smaller than the actual available bandwidth. However, in that case, as d(kT) will be smaller than $d_{ref}$, the window size will increase, causing an increase in the sending rate and thus filling the queue and providing meaningful estimation of U(kT).

## 6. EXPERIMENTATION METHODOLOGY AND EVALUATION

In order to evaluate our proposed system we performed simulations with Opnet [30] and we implemented and evaluated a real prototype under a variety of scenarios. The development of our real prototype was facilitated by experimentation and monitoring tools which have been created by BonFIRE test-bed [10]. Our experimentation scenarios have been set up by using the infrastructure of iMinds (Virtuall Wall [11]). Virtual Wall is a test bed in which set of nodes are connected through a virtual network. It gives to the test bed user the capability to create the desired network topology and dynamically adjust (during the experiments) features of each underlying network link as: path latency, path bandwidth, packet loss rate, etc. In order to evaluate the proposed congestion control architecture a network topology has been created in which a node has been assigned to act as a P2P traffic sender and a set of nodes have been used as P2P traffic receivers. Another node has been used as router in order to be able to adjust the features of each network path and to act as the bottleneck network node.

Three sets of experiments were performed in order to prove the properties of the proposed P2P congestion control. In the first one its robustness to changes in the latency of the underling network path is demonstrated. In the second is demonstrated its ability to dynamically adapt to very curt changes bandwidth of the bottleneck network point and in the third its friendliness (co-existence) to unrelated TCP traffic.

### 6.1. Path latency variation

The purpose of the first experiment is to demonstrate the robustness of the proposed architecture to the dynamic changes in the latency of the underlying network paths. In order to achieve this was created a network topology with two types of network links. The first one is the Sender-Router (bottleneck network point) link and its latency is set to a constant value equal to 20ms, while the second one is the Router-Receiver link whose latency changes every 10s according to a uniform distribution between 2ms and 22ms. The available upload bandwidth is constant and equal to 4000 Kbps. In this scenario a sender peer sends only to one receiver.

In Figure 3 we depict three variables. The first is Path minimum Round Trip Time, which changes dynamically during the execution of the experiment, the second is the desired Round Trip Time, including the time that the packets remained in the queue, and the third represents the actual measured delay between the transmission of a packet and its acknowledgment. From Figure 3 we can see that although Path RTT changes very dynamically our architecture stabilizes the queue size in the bottleneck network node at a value very close to RTTref (dREF+RTT) without ever exceeding this value. Additionally RTT average is always higher than Path RTT and never falls so low. These two observations testify that despite the changes in the underlying network path latency, our architecture fully exploits the available bandwidth without causing packet loss.

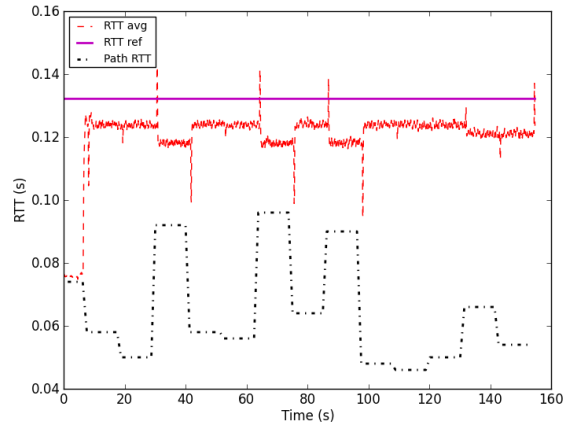

Figure 3. Path minimum Round Trip Time (Path RTT), desired Round Trip Time (RTT ref) and actual Round Trip Time (RTT avg) over time

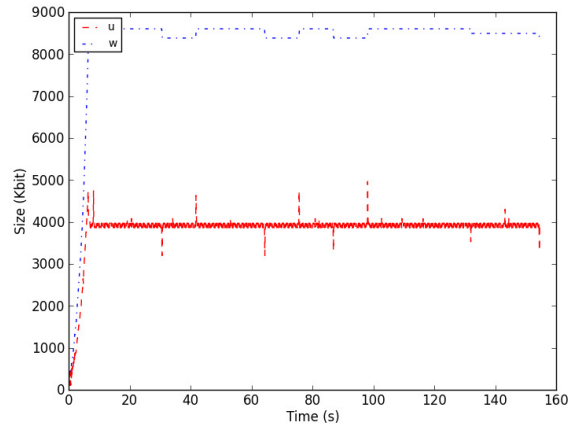

Figure 4. Amount of data to be sent, u(kT), and window, w(kT), over time

Figure 4 depicts u(kT) (as it is dynamically calculated from Eq. 1 in each iteration of the algorithm) along with the w(kT), which is the window that is calculated according to Eq. 17. Both are multiplied by the packet size in order to be translated from number of packets to Kbits. As Figure 4 shows, the window increases in the beginning when the measured RTT values are small and then it remains stable despite the variation of path latency. The spikes in u(kT) testify the immediate adaptation of our architecture to the changes of path latency.

In Figure 5 are depicted three variables. The first is the a priori set available bandwidth in the bottleneck network point (available BW). The second is the acknowledgement rate, which is dynamically measured (ack rate). The third is the calculated upload bandwidth by the sender (U).

Figure 5 demonstrates the ability of the proposed algorithm to calculate with very high accuracy the available upload bandwidth.

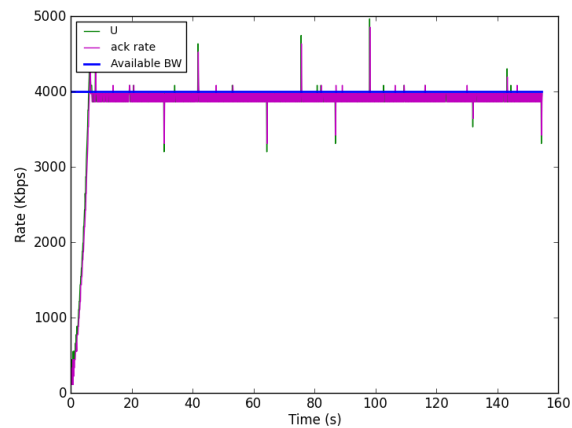

Figure 5. A priori set available bandwidth, acknowledgement rate and measured bandwidth over time

## 6.2. Path bandwidth variation

In the second set of experiments a sender sequentially sends P2P blocks to multiple receivers using the network topology of the previous experiment. The difference now is that there are four receivers connected with four Router-receiver links. The latency of each of these paths is now configured to the following values: d1=12ms, d2=22ms, d3=7ms and d4=16ms.

For better presentation of the results we define the variable dRTTi, which represents the difference between the measured latency between the sender and receiver i and the actual RTT between the sender and i (which is the RTT that was set during the deployment of the network topology). This variable represents the time interval that the packets, which were transmitted to node i, remained in the bottleneck queue and ideally should be the same with the value of the dref control variable.

In the first scenario the upload bandwidth of the sender remains constant and is set to 4Mbps. Figure 6 shows that the dRTT values for all the receivers are similar and very close to the dref value, although their respective RTT values are very different. Figure 6 testifies the ability of the proposed algorithm to control the size of the queue to the preference point during sequential transmission of P2P block to different network locations.

In the next scenario, which is represented in Figure7, Figure 8 and Figure 9, the available bandwidth in the router, that acts as the bottleneck network point, changes dynamically every 10 seconds according to a uniform distribution between 1 and 5 Mbps.

Figure 7 depicts how the available bandwidth changes over time and how the proposed P2P congestion control manages to measure the available upload bandwidth by measuring the rate of the acknowledgment's reception.

Figure 8 depicts u(kT) and w(kT) measured in Kbits. It is evident, by inspecting these two figures, that the proposed P2P congestion control is able to adapt very quickly to these very abrupt changes of the available bandwidth and to fully exploit it in every time instant.

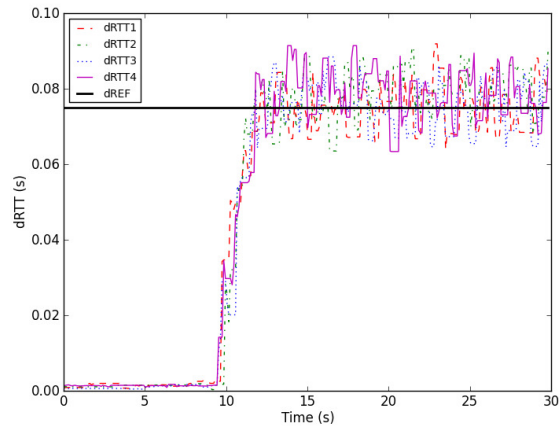

Figure 6. Queue delay measured through acknowledgements from four different receivers

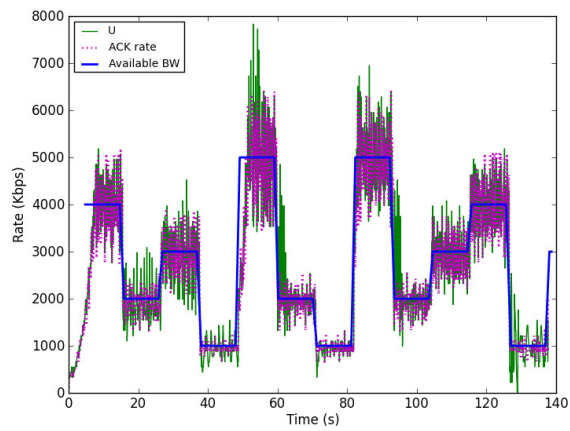

Figure 7. Available bandwidth, Acknowledge rate and measured bandwidth over time

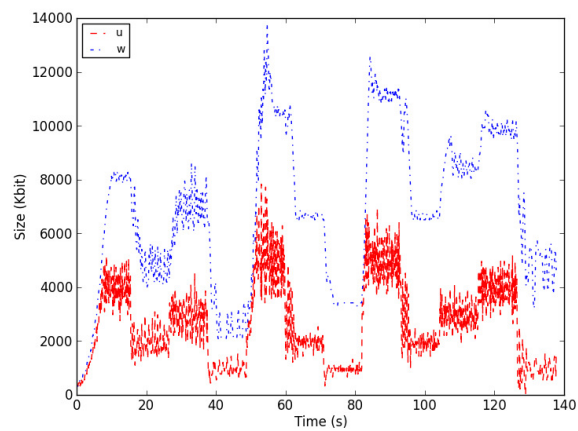

Figure 8. Amount of data to be sent, u(kT), and window, w[kT], over time

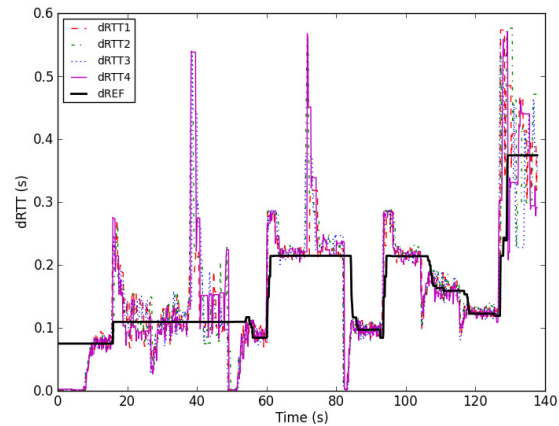

Figure 9. Bottleneck point queue delay as it is dynamically measured from acknowledgements from four different receivers

Finally, Figure 9 depicts the dRTT values for the different receivers along with the dref control variable. The interesting thing about this figure, besides the fact that all the dRTT values follow the behaviour of the $d_{ref}$, is the behavior of the $d_{ref}$ variable itself. Although the different actual RTT values remain the same for all the receivers the $d_{ref}$ variable changes over time. This is due to the recalculation over time of the $d_{pmax}$ variable (Eq. 16). In the beginning there are no errors so the value of $d_{pmax}$ is calculated by adding a constant to the $d_{pmin}$ variable. However, as the available bandwidth changes, and more specifically when it drops, packet loss occurs. This triggers the dynamic recalculation of the $d_{pmax}$ variable and as the result the recalculation of the $d_{ref}$. This dynamic behaviour of $d_{ref}$ along with its advantages will be shown more clearly in the next set of experiments

## 6.3. Co-existence with TCP

The third set of experiments analyses the behaviour of the proposed P2P congestion control under the existence of TCP traffic. In order to take meaningful results two different congestion control algorithms of TCP were used. These are TCP-BIC [15] and TCP-RENO [16]. The setup for the remaining experiments is the same with the second set, having four receivers with fixed path latencies according to the previous distribution. The capacity of the bottleneck node is fixed and set to 4Mbps. The duration of the experiments is 100s and by using iperf [17] TCP data is sent to receiver 1 in parallel with the P2P flows. The value of the constant α in Eq. 16 is set to 0.75.

**TCP-BIC**: In the first experiment, where TCP-BIC coexists with the proposed P2P congestion control, TCP flow starts after our algorithm has been running for 40s and lasts for 60s. Figure 10, Figure 11 and Figure 12 are the same three graphs that we presented for the first two set of experiments .Figure 10 depicts the recalibration of the $d_{ref}$ variable. When TCP traffic starts at time 40 the queue gets full and an error occur. This triggers the right calculation of the $d_{pmax}$ and thus of the $d_{ref}$. From that point on $d_{ref}$ remains constant, meaning there are no more lost packets, and the congestion control algorithm succeeds in keeping the dRTT values close to the desired point. Figure 11 and Figure 12 show how the control algorithm adapts to the presence of the TCP traffic. The u(kT) and w(kT) drops and stabilize in time, while quickly regaining their initial value after the termination of the TCP traffic. The above experiment depicts the ability of the proposed P2P congestion control to not starve and to stably continue to send data despite TCP-BIC trying to push the queue latency (dRTT) to the desired value.

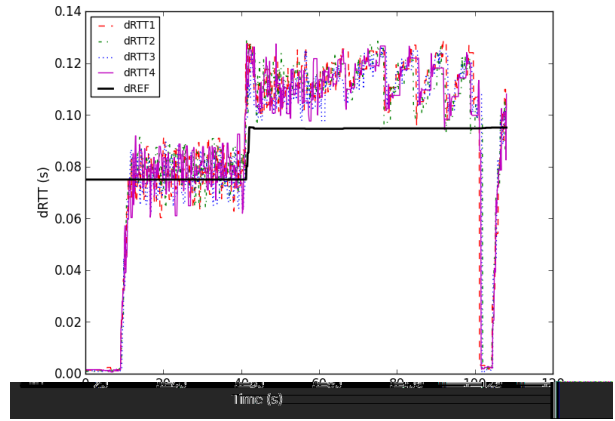

Figure 10. Queue delay measured through acknowledgements from four different receivers

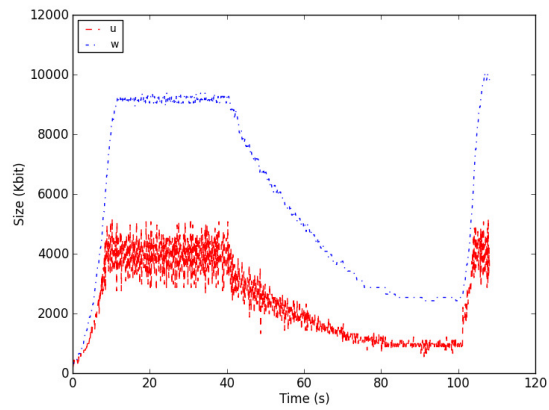

Figure 11. Amount of data to be sent, u(kT), and window, w[kT], over time

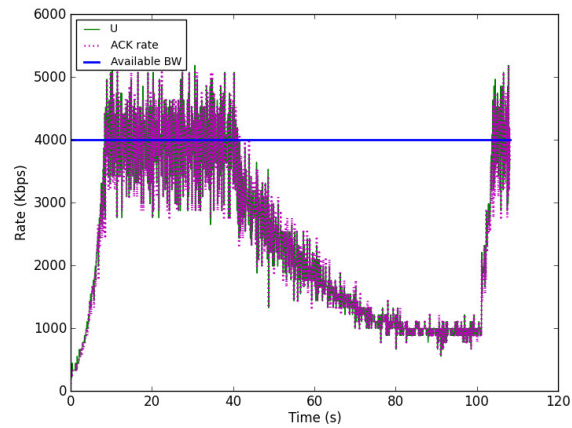

Figure 12. A priori set available bandwidth, Acknowledge rate and measured bandwidth over time

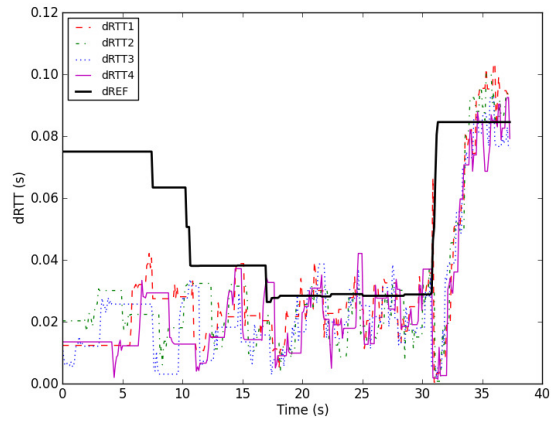

Figure 13. Queue delay as it is measured from acknowledgements from four different receivers

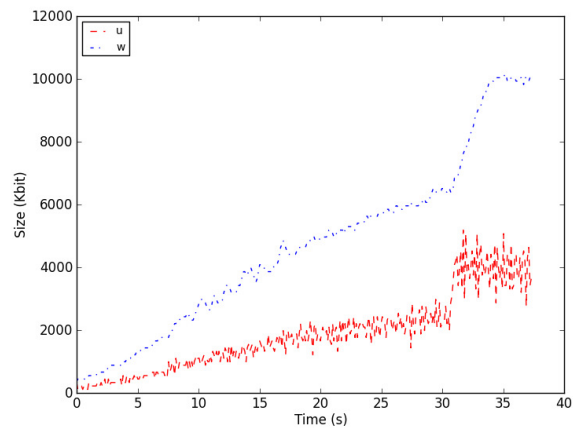

Figure 14. Amount of data to be sent, u(kT), and window, w[kT], over time

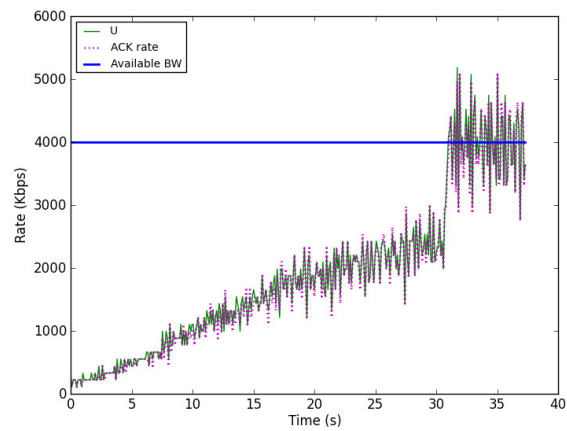

Figure 15. A priori set available bandwidth, Acknowledge rate and measured

In the second experiment with TCP-BIC which is depicted in Figure 13, Figure 14 and Figure 15 the proposed P2P congestion control architecture is started after the TCP has been running for 30s (time 0). We can observe that, also in this scenario, the proposed congestion control manages to "compete" fair with the TCP traffic and allocate half the available bandwidth (2 Mbps), while, when the TCP flow ends (time 30s), it quickly adjusts and uses all the available which is 4Mbps

**TCP-RENO**: The TCP-RENO experiment is identical with TCP-BIC experiment. Figure 16, Figure 17 and Figure 18 depict the case that TCP traffic started at 15 sec and ended at 75sec. As TCP-RENO doesn't behave as aggressively as TCP-BIC, the allocated bandwidth of the P2P congestion control quickly converges at 2Mbps, which is half the capacity of the link.

In the second TCP-RENO experiment, P2P traffic started after the TCP stream has been running for 25s. In Figure 19, Figure 20 and Figure 21 we see how P2P congestion control tends to converge around 3Mbps.

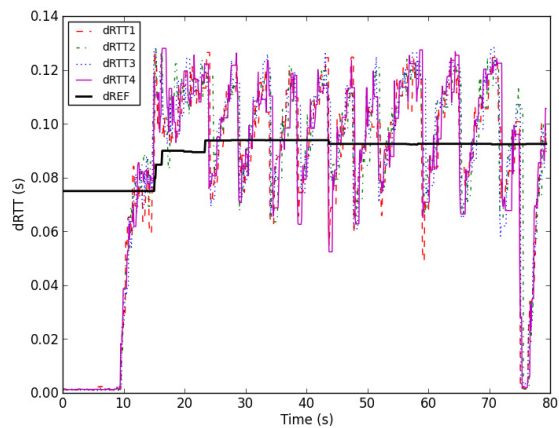

Figure 16. Bottleneck point queue delay as it is dynamically measured from acknowledgements from four different receivers

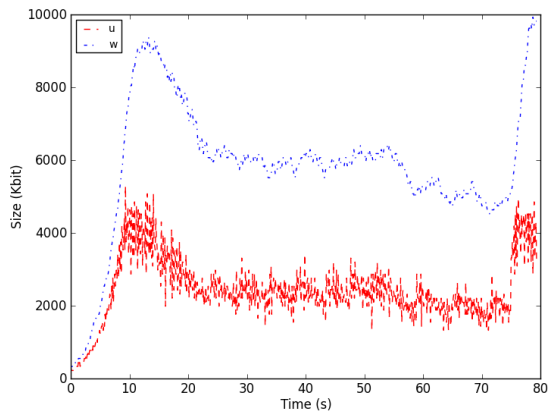

Figure 17. Amount of data to be sent, u(kT), and window, w[kT], over time.

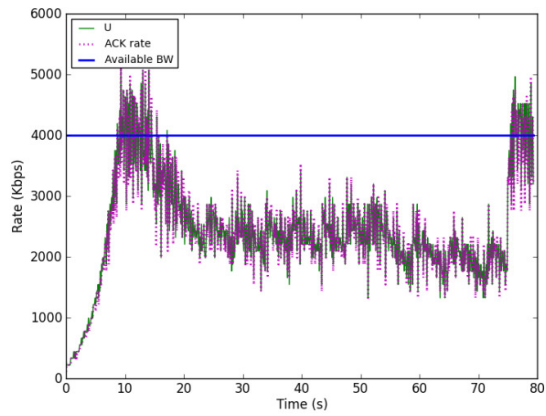

Figure 18. A priori set available bandwidth, Acknowledge rate and measured bandwidth over time

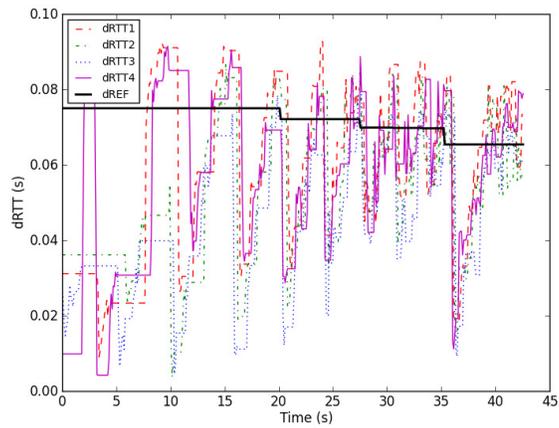

Figure 19. Queue delay as it is measured from acknowledgements from four different receivers

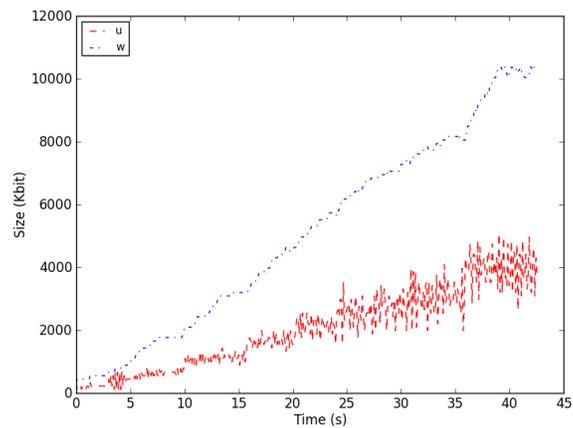

Figure 20. Amount of data to be sent, u(kT), and window, w[kT], over time.

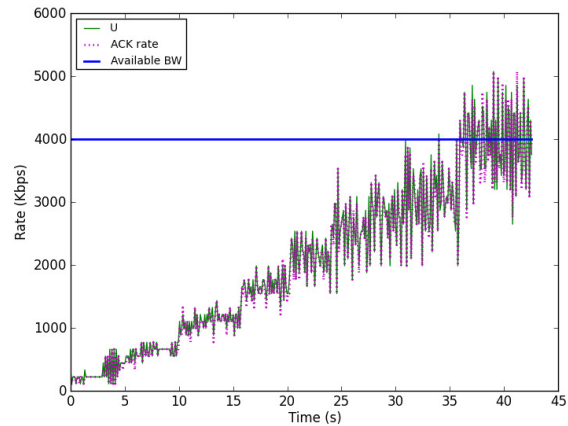

Figure 21. Available bandwidth, Acknowledge rate and measured bandwidth over time

## 6. CONCLUSIONS

Traditional congestion control algorithms are designed for bulk point to point transfers. In this work we designed, implemented and evaluated a P2P congestion control algorithm suitable for flows containing small chunks of data which are transmitted sequentially to different network destinations. Our theoretical work was justified though our evaluation in a real and implemented system and we proved that our proposed P2P congestion control architecture: i) is able to utilize efficiently all the upload bandwidth of participating peers, ii) is stable, robust even under sudden and large changes in the bandwidth of the bottleneck network point, iii) is immune to time-varying delays (underlying path latency) caused by dynamic underlying network traffic, iv) is able to measure accurately and dynamically the upload bandwidth capacity of each peer, v) is able, by controlling the queue length in the bottleneck network point, to avoid buffer overloading in the Home Gateways and routers of the underlying network.

## ACKNOWLEDGEMENTS

This work was funded from BonFIRE [10] which is an EU project funded by the EC FP7 under grant agreement number 257386.

## AUTHORS

**Nikolaos Efthymiopoulos** received the diploma and Doctor of Philosophy degree in Electrical and Computer Engineering from the University of Patras, Greece, in 2004 and 2010, respectively. His main research interests are: network optimization, network control, scalable systems, peer to peer, distributed video streaming, distributed searching and achieving QoS in computer networks. He has more than 20 publications in these areas. He more than 10 years of experience in several FP7 ICT projects and he was technical manager assistant and WP leader in VITAL++ and STEER. He has temporarily worked as an Assistant Professor in Informatics & MM Department in Greece. He is currently a Post-Doctoral Research Associate at the University of Patras in Greece.

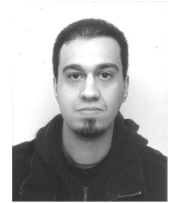

**Athanasios Christakidis** received his diploma in 2004 and his Doctor of Philosophy in 2010 from the Department of Electrical and Computer Engineering at the University of Patras, Greece. His research interests are peer to peer networks, distributed optimization, network resource allocation, and congestion control. Since 2004, he participated in several FP7 projects, and he has more than 15 publications in these areas. He has led the development of a client for stable and efficient peer to peer live streaming.

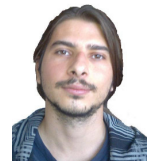

**Maria Efthymiopoulou** received the first degree and the Doctor of Philosophy degree in Electrical and Computer Engineering from the University of Patras, Greece, in 2008 and 2015, respectively. Her main research interests are: network control, scalable systems, peer to peer, live streaming, video on demand, development of simulation environments, QoS in computer networks. She has several publications in these areas. She has 7 years of experience in several FP7 ICT projects. She is currently a Post-Doctoral Research Associate at the University of Patras in Greece.

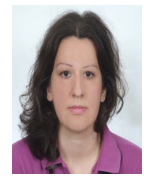

**Loris Corazza** was born in Italy on 1982. He approached to University of Patras as member of the Network Architectures and Management group in 2009. His involvement in ICT projects started in 2008, when he was member of the Smart Systems Team in Hitachi SAS Sophia-Antipolis Research Lab. He has been one of the head developers of P2NER P2P-Client while researching in the area of P2P Systems and Content Distribution Networks. He is currently a researcher at University of Patras, Greece in the area of Optical Software Defined Networks. His main interests and expertise are: software design and architecture, network protocols, network management, algorithms for data distribution in P2P networks and performance evaluation, network security.

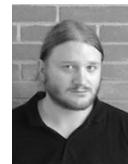

**Spyros Denazis** is a Professor in the Electrical & Computer Engineering Department, at the University of Patras, Greece. He received his Doctor of Philosophy in Computer Science from the University of Bradford, UK, in 1993. In 1996, he joined the R&D Department of Intracom SA in Athens as Project Coordinator, and in 1998, he joined the Information Technology Laboratory of Hitachi Europe in Cambridge UK, as a Senior Research Engineer while serving for 3 years (1998-2001) as an Industrial Research Fellow in the Centre for Communications Systems Research , of Cambridge University, UK. For the period 2003-2010, he had also been a Consultant for the Hitachi Europe Sophia Antipolis Laboratory, in France. Currently, he leads the Network Architecture & Management Group where he coordinates a range of research activities in the areas of P2P live streaming, future internet research experimentation, and SDN.

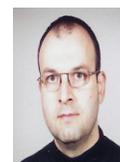